\documentclass[a4paper,12pt]{JHEP3}
\newcommand{\be}{\begin{equation}}
\newcommand{\ee}{\end{equation}}
\newcommand{\bq}{\begin{eqnarray}}
\newcommand{\eq}{\end{eqnarray}}

\newcommand{\one}{\hbox{\rm 1\kern-.27em I}}

\title{Faraday's Lines of Force as Strings: from Gauss' Law to the Arrow of Time} 
\author{Paul Mansfield \\Centre for Particle Theory, University of
Durham, Durham DH1 3LE, UK \\ Email: P.R.W.Mansfield@durham.ac.uk}
\abstract{We reformulate classical electromagnetism as the statistical mechanics of lines of electric flux with dynamics described by the string action in four dimensions. The retarded solution to Maxwell's equations emerges naturally as an average over a microcanonical ensemble of these lines of force at high temperature.}


\preprint{DCPT-11/29}

\begin{document}

\section{\bf Introduction}
Faraday considered lines of force to be a physical substance, the basic dynamical object of electromagnetism, and indeed other physical theories. In his view\footnote{``You are aware of the speculation which I sometime since uttered respecting the view of the nature of matter which considers its ultimate atoms as centres of force, and not as so many little bodies surrounded by forces....The view which I am so bold as to put forth considers, therefore, radiation as a high species of vibration in the lines of force....It endeavours to dismiss the ether, but not the vibrations.'' \cite{Faraday}} the particles acted on by forces were not separate entities but actually configurations of forces, and the lines themselves
could be physically shaken, disturbances propagating along them with finite speed thus 
accounting for radiation without the need of an ether.

 As the quantitative description of electromagnetism was developed this point of view lost ground to Maxwell's which
took the electric and magnetic fields, as well as moving charges, as the dynamical degrees of freedom \cite{Maxwell}.
Gauss' law\footnote{We will use S.I. units to express Maxwell's equations \cite{Pan}} in integral form $\oint_\Sigma {\bf E}\cdot d{\bf S}=\int \rho\,dV/\epsilon_0$ seems to support Faraday's interpretation. It measures the number of lines of electric force cutting a closed surface, $\Sigma$, as though these were physical objects capable of being counted and sets it equal to the enclosed charge. In differential form, $\nabla\cdot{\bf E}=\rho/\epsilon_0$, this becomes just a differential equation that relates ${\bf E}$ and $\rho$, on the same footing as the other Maxwell equations that together determine the dynamics of the theory. From Maxwell's point of view the lines of force are no more than a geometric representation of the field, and the integral version of Gauss' law simply a statement about the ends of the curves used in that representation. 

The canonical quantisation of electrodynamics moves even further from Faraday's picture. Based as it is on a Hamiltonian formulation, the dynamical degrees of freedom are the gauge potentials $A_0, \,{\bf A}$, modulo gauge tranformations, rather than $\bf E$ and $\bf B$. Their existence follows from the Maxwell equations $\nabla\cdot {\bf B}=0$ and $\nabla\times{\bf E}=-\dot{\bf B}$, which are therefore implemented as identities in the quantum theory. The remaining Maxwell equations, i.e. Gauss' law and 
\be
\nabla\times {\bf B}=\mu_0 {\bf J}+\mu_0\epsilon_0\dot{\bf E}
\label{amp}
\ee
are the Euler-Lagrange
equations for the theory. To quantise one has to pick a gauge.
With the choice $A_0=0$, which is most convenient for the Hamiltonian formalism,  (\ref{amp}) is  Hamilton's equation of motion. However the choice of gauge removes $A_0$ as a dynamical variable so that Gauss'
law is not recovered in this way, and has to be imposed as a constraint. In the quantum theory this becomes a restriction on the physical states of the theory. The interpretation of $\nabla\cdot{\bf E}$ as the generator of the remaining gauge transformations that preserve the gauge condition implies that the physical states are selected to be those that are invariant under time-independent gauge transformations. So Gauss' law, which from Faraday's point of view counts the 
dynamical objects in the theory, is not even valid for all the states needed to construct the quantum theory, but only for the physical subspace.

There is also a conceptual difficulty with treating lines of force as dynamical objects. The lines, being tangent to the field, encode its direction. Its magnitude is represented by their density. As fields vary continuously with position they cannot be modeled by a whole number of lines, making obscure the concept of individual lines as dynamical objects. 

The purpose of this paper is to attempt to resurrect, albeit in modified form, the notion that lines of force can indeed be treated 
as dynamical objects in their own right. In classical electromagnetism the lines of force are fixed by the charge distribution, but we will consider the consequences of allowing their positions to fluctuate.
We will overcome the conceptual difficulty by assuming that the large number of lines of force stretching between macroscopic charges should be treated using statistical mechanics, so that the classical electromagnetic field emerges as an average using an appropriate Boltzmann weight, and thus can vary continuously with position. String theory provides us with a natural identification of this weight and the technology to compute the average.

In 1955 Dirac proposed a similar solution to this conceptual difficulty \cite{Dirac}. He was interested in constructing a version of QED in which the electrons were created by gauge invariant operators which create a part of the electromagnetic field along with the charge. By taking this to consist of a single Faraday line of force associated with a quantum of flux he provided another explanation of electric charge quantisation (in addition to his celebrated argument based on the motion of an electric charge in the presence of a magnetic monopole.) In this theory closed lines of force describe photons, open lines describe electron positron pairs and pair creation is described by the breaking of lines of force. Although the flux is supported on the line of force quantum superposition allows for states with continuous fields, such as the spherically symmetric field of a single charge. In our approach we will construct the classical electromagnetic field as a thermal average over a macroscopic number of lines of force connecting electric charges that are large in comparison with the charge  of the electron. Dirac's theory could be taken as the microscopic description underlying this.

We will begin by studying two simplified cases namely electrosatics and magnetostatics. The first is easy to formulate mathematically, whilst the second already requires the formalism of string theory and is a useful stepping stone to constructing the full time-dependent electromagnetic field.
By magnetostatics we mean the time independent magnetic field generated by constant currents flowing around closed circuits.
The lines of force associated with ${\bf B}$ form closed curves because $\nabla\cdot {\bf B}=0$,
and Farady thought of these as dynamical objects too. However,
for this particular example we will not focus on these lines of force. Instead we will take the dynamical objects that describe magnetism to be
surfaces spanning the circuits the currents flow round. Again there is a natural weight to average over these surfaces, but we might expect to be impeded by  the well-known difficulties encountered in trying to formulate sums over random surfaces that make it difficult to construct
string theory away from its critical dimension \cite{Ginsparg:1993is}. Remarkably these difficulties are absent from our problem even though it amounts to an off-shell calculation in non-critical string theory and we are able to evaluate the sum and show that it yields the Biot-Savart law. 

The magnetostatic problem generalises to higher dimensions in which context the fluctuating surfaces can be re-interpreted as the world-sheets of lines of force. By choosing the target-space appropriately we will show that the retarded solution to Maxwell's equations arises naturally as a thermal average over these lines, so that by invoking statistical mechanics this approach violates the usual time-reversal invariance of classical electromagnetism.

\section{Electrostatics}
We begin with the special case of a static electric field.
Consider the field, ${\bf E}({\bf x})$, due to two equal and opposite charges, $\pm q$, placed at ${\bf a}$ and ${\bf b}$ respectively. 
\be
{\bf E}({\bf x})={q\over 4\pi\epsilon_0}\,{{\bf x}-{\bf a}\over ||{\bf x}-{\bf a}||^3}-
{q\over 4\pi\epsilon_0}\,{{\bf x}-{\bf b}\over ||{\bf x}-{\bf b}||^3}\,.
\label{E}
\ee
This is the unique solution to 
$\nabla\times {\bf E}=0$ and
the differential form of Gauss' law
\be
\nabla\cdot {\bf E}({\bf x})={q\over\epsilon_0}\,\delta^3( {\bf x}-{\bf a})-{q\over\epsilon_0}\,\delta^3( {\bf x}-{\bf b})\,,\label{dGauss}
\ee
that decays at infinity.
Briefly setting aside the former of these two differential equations, it is easy to see that 
(\ref{dGauss}) is solved by taking the electric field to be given by
\be
{\bf E}'({\bf x})={q\over\epsilon_0}\int_C \delta^3( {\bf x}-{\bf y}) \,d{\bf y}\label{DiracString}
\ee
for any curve $C$ from ${\bf a}$ to ${\bf b}$, since for any differentiable test-function $u({\bf x})$
that vanishes at infinity:
$$\int \nabla\cdot {\bf E}'({\bf x})\,u({\bf x})\,dV=-\int {\bf E}'({\bf x})\cdot\nabla u({\bf x})\,dV=-\int    {q\over\epsilon_0}\left(\int_C \delta^3( {\bf x}-{\bf y}) \,d{\bf y}\right)\cdot     \,\nabla u({\bf x})\,dV$$
\be
=-{q\over\epsilon_0} \int_C \nabla u({\bf y})\cdot d{\bf y} ={q\over\epsilon_0}\,u({\bf a})-{q\over\epsilon_0}\,u({\bf b})=\int {q\over\epsilon_0}\,\left(\delta^3( {\bf x}-{\bf a})-\,\delta^3( {\bf x}-{\bf b})\right)\,u({\bf x})\,dV\,. \label{calc}
\ee
(\ref{DiracString}) has the same mathematical form as the Dirac string used to represent the magnetic field of a monopole. In that context the position, $C$, of the string has no physical significance, and can be changed by a particular kind of gauge transformation. In our work, as in \cite{Dirac}, however, we will attach physical meaning to $C$, treating it as the position of a physical object. Now the electric field of 
(\ref{DiracString}) is supported on $C$
and so is completely different in character to that of (\ref{E}) which is supported everywhere.
We could describe the physical object that has position $C$ as a string of electric flux.
To obtain (\ref{E}) we will make the assumption that the theory is stochastic in the sense that the positions of the flux strings are to be averaged over with a Boltzmann weight, $e^{-\beta {\cal{H}}}$, so that
${\bf E}({\bf x})=\langle {\bf E}'({\bf x})\rangle_C$. The average of any functional
of $C$, $\Omega$ is given by the functional integral
$$
\langle \Omega\rangle_C={1\over Z}\int {\cal D} {\bf y}\,\Omega\,e^{-\beta {\cal H}[{\bf y}]}\,,
$$
(with $Z$ a normalisation constant, so that $\langle  1\rangle_C=1$). The physical interpretation is that the macroscopic charge $q$ generating the classical electric field is composed of many microscopic or elementary charges, of magnitude $q_0$, each of which is the terminus of a line of force as in Dirac's theory. These lines are physical objects in thermodynamic equilibrium at temperature $1/\beta$ and there are $q/q_0$ of them. Each has an electric field given by (\ref{DiracString}) with $q$ replaced by $q_0$, and these contributions add up to give the total electric field.

We keep the end-points of all the curves to be averaged over fixed at $\bf a$ and $\bf b$ so the
calculation (\ref{calc}) goes through as before for the averaged field and Gauss' law is satisfied.
We now have to find $\beta {\cal H}$ so that $\nabla \times {\bf E}=0$, or equivalently, so that
\be
{{\bf x}-{\bf a}\over ||{\bf x}-{\bf a}||^3}-
{{\bf x}-{\bf b}\over ||{\bf x}-{\bf b}||^3}={1\over Z}\int {\cal D} {\bf y}\,\int_C \delta^3( {\bf x}-{\bf y}) \,d{\bf y}\,e^{-\beta {\cal H}[{\bf y}]}\, .
\label{Ee}
\ee
There is a natural choice for $\beta {\cal H}$ that occurs in the path-integral representation of quantum mechanics \cite{Feynman}
and the heat-kernel connected with diffusion and Brownian motion. If ${\bf y}(t)$, $0\le t\le T$ is a parametrisation of a path from ${\bf a }$ to ${\bf b}$ then
\be
\langle{\bf b}|e^{-T \hat H_0}| {\bf a}\rangle=\int {\cal D} {\bf y}\,e^{-\int_0^T dt\,\dot{\bf y}^2/2}={e^{-||{\bf a}-{\bf b}||^2/(2T)}\over (2\pi T)^{3/2}}\label{evol}
\ee
where $\hat H_0=\hat{\bf p}^2/2$ so $2\langle{\bf b}|\hat H_0={\nabla^2}\langle{\bf b}|$ and the eigenstates of position are normalised
to $\langle{\bf b}|{\bf a}\rangle=\delta^3({\bf a}-{\bf b})$. If we take $\beta=1/T$ and ${\cal H}=\int_0^1 du\, (d{\bf y}/du)^2/2$ then by a change of variable, $t=Tu$, we get $\beta {\cal H}=\int_0^T dt\,\dot{\bf y}^2/2$. We have to set a value to $T$. This is a dimensionful quantity, and no such parameter appears in (\ref{Ee}) 
so we will take the limit in which $T$ is large (in comparison to the squares of the other lengths in our problem.) The expectation value of the delta-function can be generated by functionally differentiating with respect to a source term added to $\beta {\cal H}$:
\be
\int {\cal D} {\bf y}\,\int_C \delta^3( {\bf x}-{\bf y}) \,d{\bf y}\,e^{-\int_0^T dt\,\dot{\bf y}^2/2}=\left\{{\delta\over\delta {\bf \cal A}({\bf x})}
\int {\cal D} {\bf y}\,e^{-\int_0^T dt\,\dot{\bf y}^2/2 + \int_{\bf a}^{\bf b} {\bf \cal A}({\bf y})\cdot d {\bf y}}
\right\}\Big |_{{\bf \cal A}=0}
\label{Eee}
\ee
The functional integral inside the braces is the generalisation of (\ref{evol}) to a particle moving in an electro-magnetic field with vector potential $i{\bf \cal A}$, so the Hamiltonian $\hat H_0$ is modified to $\hat H=(\hat{\bf p}+i{\bf \cal A})^2/2$. Thus (\ref{Eee}) can be written as
$$
{\delta\over\delta {\bf \cal A}({\bf x})}
\langle{\bf b}|e^{-T\hat H}| {\bf a}\rangle\Big |_{{\bf \cal A}=0}
=-\int_0^T dt\,\langle{\bf b}|e^{(t-T)\hat H_0}{\delta \hat H \over\delta {\bf \cal A}({\bf x})}\Big |_{{\bf \cal A}=0}e^{-t\hat H_0}
| {\bf a}\rangle
$$
Now when $ {{\bf \cal A}=0}$,
${2\,\delta \hat H /\delta {\bf \cal A}({\bf x})}=
i\hat {\bf p} \,\delta^3(\hat {\bf q}-{\bf x})+
\delta^3(\hat {\bf q}-{\bf x})\,i\hat {\bf p}$, so 
using the resolution of the identity $\int | {\bf c}\rangle \,d^3{\bf c}\,\langle{\bf c}|=\one$,
gives 
$$2{\delta \hat H \over\delta {\bf \cal A}({\bf x})}\Big |_{{\bf \cal A}=0}=-\int
\left(\nabla_{\bf c}\,| {\bf c}\rangle \right )\,d^3{\bf c}\,\langle{\bf c}|\,\delta^3({\bf c}-{\bf x})
+\int\delta^3({\bf c}-{\bf x})\,| {\bf c}\rangle \,d^3{\bf c}\,{\nabla}_{\bf c}\langle{\bf c}|
= | {\bf x}\rangle \,\stackrel{\leftrightarrow}{\nabla}\,\langle{\bf x}|$$
so that (\ref{Eee}) becomes (after setting
$ {{\bf \cal A}=0}$)
\be
-{1\over 2}\int_0^T dt\,\langle{\bf b}|e^{(t-T)\hat H_0}
| {\bf x}\rangle \,\stackrel{\leftrightarrow}{\nabla}\,\langle{\bf x}|
e^{-t\hat H_0}
| {\bf a}\rangle\,.
\ee
The normalisation constant, $Z$ is just the right-hand-side of (\ref{evol}) so now we have 
$$
{1\over Z}\int {\cal D} {\bf y}\,\int_C \delta^3( {\bf x}-{\bf y}) \,d{\bf y}\,e^{-\beta{\cal H}({\bf y})}
=-{(2\pi T)^{3/2}\over2e^{-||{\bf a}-{\bf b}||^2\over2T}} \int_0^T dt\,{e^{-||{\bf x}-{\bf b}||^2\over2(T-t)}\over (2\pi (T-t))^{3/2}}
\,\stackrel{\leftrightarrow}{\nabla}\,
{e^{-||{\bf a}-{\bf x}||^2\over 2t}\over (2\pi t)^{3/2}}\,.
$$
For large $T$ the integrand is negligible except when 
$t\approx 0$ and $t\approx T$, so that in the limit of infinite $T$ the integral separates into two contributions:
$$-\int_0^\infty dt\,\nabla {e^{-||{\bf a}-{\bf x}||^2/(2t)}\over 2(2\pi t)^{3/2}}+
\int_0^\infty dt\,\nabla {e^{-||{\bf x}-{\bf b}||^2/(2t)}\over 2(2\pi t)^{3/2}}=\nabla\left(
-{1\over 4\pi||{\bf x}-{\bf a}||}+{1\over 4\pi||{\bf x}-{\bf b}||}\right)
$$
which yields the right-hand-side of (\ref{Ee}). 

When we add up the contributions to the electric field of the 
individual lines of force we obtain
\be
\langle \,{q\over\epsilon_0}\int_C \delta^3( {\bf x}-{\bf y}) \,d{\bf y}\,\rangle_C={q\over 4\pi\epsilon_0}\,{{\bf x}-{\bf a}\over ||{\bf x}-{\bf a}||^3}-
{q\over 4\pi\epsilon_0}\,{{\bf x}-{\bf b}\over ||{\bf x}-{\bf b}||^3}={\bf E}({\bf x})\label{exp}\,.
\ee

\section{Magnetostatics}

We now consider another simplified case, namely the magnetic field generated by a current that is constant in time. Although this is a different physical problem to that of the preceding section the mathematical description will provide a stepping-stone to introducing time-evolution into the description of the electromagnetic field generated by point charges. A constant current, $I$, flowing around a circuit $C$ has density
\be
{\bf J}({\bf x})=I\oint_C \delta^3( {\bf x}-{\bf y}) \,d{\bf y}\,.\label{current}
\ee
$\nabla\cdot{\bf J}=0$ follows from a similar argument to (\ref{calc}). The constant magnetic field it generates is given by the Biot-Savart law
\be
{\bf B}({\bf x})={\mu_0I \over 4\pi}\oint_C {d{\bf y}\times(
{\bf x}-{\bf y})\over ||{\bf x}-{\bf y}||^3}\,,
\label{bs}
\ee
which is the unique solution to the Maxwell equations $\nabla\cdot {\bf B}=0$ and $\nabla\times {\bf B}=\mu_0
{\bf J}$ that vanishes at infinity. The first of these implies that the lines of magnetic force are closed, however we will not focus on the flux-lines, but consider the representation of ${\bf B}$ in terms of surfaces normal to it with density proportional to its magnitude, i.e. the equipotential surfaces for the (multi-valued) magnetic scalar potential proportional to the solid angle subtended by $C$ at the point $\bf x$. In integral form the second of this pair of Maxwell equations is Amp\`ere's law
\be
\oint_{C'} {\bf B}\cdot d{\bf x}=\int_{\Sigma'}\mu_0
{\bf J}\cdot d{\bf S}\,.
\ee
If $C'$ is chosen to loop around $C$ then the right-hand-side of this is $\mu I$ and the left-hand-side counts the number of surfaces representing ${\bf B}$ that are cut by $C'$.
These open surfaces all have boundary $C$.
This is a generalisation of Gauss' law, and we will interpret it similarly as an indication that the theory can be rewritten in terms of dynamical objects which are the surfaces spanning $C$. In classical electromagnetism these surfaces are of course fixed once $C$ is specified, but we will investigate the consequences of allowing the surfaces to fluctuate so that the magnetic field is obtained by averaging over them with an appropriate weight.

$\nabla\times {\bf B}=\mu_0
{\bf J}$ is solved by taking the magnetic field to be
\be
{\bf B}'({\bf x})=\mu_0I\int_\Sigma \delta^3({\bf x}-{\bf y})
\,d{\bf S}({\bf y}) \,,\label{bprime}
\ee
where $\Sigma$ is any surface spanning the current circuit $C$. This is readily shown by integrating ${\bf B}'$ against the curl of a vector test-function. Furthermore this remains true on averaging over all such surfaces spanning $C$ with any weight. We will now endeavour to find a weight so that 
the averaged field also satisfies $\nabla\cdot {\bf B}=0$. As in the previous example of electrostatics there is a natural choice, but before we invoke it we return to that used for
averaging over the lines of electric force and write it in a form that will motivate the generalisation to surfaces. In the previous section we used
\be
\langle \Omega\rangle_C=\lim_{T\rightarrow\infty}{1\over Z}\int {\cal D} {\bf y}\,\Omega\,e^{{-\int_0^T dt\,\dot{\bf y}^2/2}}\,.\label{avv}
\ee
Consider replacing the exponent $\int_0^T dt\,\dot{\bf y}^2/2$ by \cite{Brink:1976uf}
\be
{1\over 2T}\int_0^1 {g^{-1}(\xi)}\,{d{\bf y(\xi)}\over 
d\xi}^2 \,\sqrt{g(\xi)}\,d\xi \label{bdvs}
\ee
where ${\bf y(\xi)}$, $0\le \xi\le 1$ is a different parametrisation of the path and $g(\xi)>0$ is a new variable.
(\ref{bdvs}) 
is invariant under diffeomorphisms $\xi\rightarrow\tilde\xi$ that preserve the parameter interval provided that $g(\xi)$ transforms as an intrinsic metric $g(\xi)\,d\xi^2\rightarrow \tilde g(\tilde\xi)\,d\tilde\xi^2=g(\xi)\,d\xi^2$ and
${\bf y(\xi)}\rightarrow \tilde{\bf y}(\tilde\xi)
={\bf y(\xi)}$.
To construct this new weight we have to choose some value for $g(\xi)$, but which value we choose will not affect the result for the electric field, as we will see. If we change parameter from $\xi$ to $t$ given by $t=T\int_0^\xi \sqrt{g(\xi')} \,d\xi'$ then (\ref{bdvs})
becomes $\int_0^{T'} dt\,\dot{\bf y}^2/2$ with $T'=T\int_0^1\sqrt g\,d\xi$. Using this in (\ref{avv}) 
gives the same results as $T$ and $T'$ tend to infinity.

This form of the weight has a natural generalisation to the sum over surfaces used in Polyakov's approach to the bosonic string, \cite{Polyakov}. Let a surface $\Sigma$ spanning $C$ be parametrised by ${\bf Y}(\xi^1,\xi^2)$ with the `world-sheet co-ordinates' $\xi^a$ lying in some fixed domain $D$, then
\be
\langle \Omega\rangle_\Sigma={1\over Z}\int {\cal D} {\bf Y}\,\Omega\,\exp\left({
-{1\over 4\pi\alpha'}\int_D g^{ab}{\partial{\bf Y}\over\partial\xi^a}\cdot{\partial{\bf Y}\over\partial\xi^b}\,\sqrt {g}\,d^2\xi}\right)
\,,\label{avvS}
\ee
where $g_{ab}$ plays the r\^ole of an intrinsic metric on $\Sigma$, $g^{ab}$ is its inverse, $g=\det (g_{ab})$,
and $\alpha'$ is a dimensionful constant. 
We will refrain from integrating over $g_{ab}$, but rather choose a value for it and find that, as before, the averaging does not depend on the value we pick. 
We will now show that with this weight 
\be
{\bf B}({\bf x})=\langle\,\mu_0I\int_\Sigma \delta^3({\bf x}-{\bf Y})
\,d{\bf S}({\bf Y}) \,\rangle_\Sigma\label{BB}
\ee
satisfies the Biot-Savart law (\ref{bs}). In doing so we will encounter the usual problem of trying to formulate string theory away from its critical dimension. 

We evaluate (\ref{BB}) in the standard way by first  exponentiating the $\bf Y$ dependence, using a Fourier decomposition of the delta-function and generating  
$d{\bf S}={1\over 2}\epsilon^{ab}(\partial {\bf Y}/\partial \xi^a)\times (\partial {\bf Y}/\partial \xi^b)$ by differentiation with respect to sources:
$${\bf B}({\bf x})=
\langle\,\int_\Sigma \delta^3({\bf x}-{\bf Y})
\,d{\bf S}({\bf Y}) \,\rangle_\Sigma
=\int {d^{3} k\over 32\pi^4\alpha'}\, d^2\xi\, \epsilon^{ab}{\partial\over\partial{
\bf j}^a}\times{\partial\over\partial{
\bf j}^b}\,{1\over Z}\int {\cal D} {\bf Y}\,e^{-S'}\Big |_{{\bf j}=0}\,,
$$
$$
2\pi\alpha'S'=\int_D \left(g^{ab}{1\over 2}{\partial{\bf Y}\over\partial\tilde\xi^a}\cdot{\partial{\bf Y}\over\partial\tilde\xi^b}\,\sqrt {g}\
+\left\{i{\bf k}\cdot ({\bf x}-{\bf Y})
+{\bf Y}\cdot {
\bf j}^a{\partial\over\partial\tilde\xi^a}\right\}\,\delta^3(\tilde \xi-\xi)
\right)\,d^2\xi
$$
The dependence on $\bf k$ and $\bf j$ is separated out by
writing $\bf Y$ as the sum of a classical solution to the Euler-Lagrange equations for $S'$, ${\bf Y}_c$, and a quantum fluctuation, $\bar{\bf Y}$:
$$
{\bf Y}={\bf Y}_c+\bar{\bf Y}\,,
\quad  
-{\partial\over\partial\tilde\xi^a}\left(\sqrt {g}g^{ab}{\partial{\bf Y}_c\over\partial\tilde\xi^b}\right)
=\left\{i{\bf k}
-{
\bf j}^a{\partial\over\partial\tilde\xi^a}\right\}\,\delta^3(\tilde\xi-\xi)
$$
where on the boundary of $D$, ${\bf Y}_c$ coincides with the current circuit $C$, i.e. $\bf y$,
 and $\bar{\bf Y}$ vanish.
${\bf Y}_c$ can be found using the Dirichlet Green function for the Laplacian on $D$, $G$,
$$
{\bf Y}_c(\tilde\xi)=\int_DG(\tilde\xi,\xi')
\left\{i{\bf k}
-{
\bf j}^a{\partial\over\partial{\xi'}^a}\right\}\,\delta^3(\xi'-\xi)\,d^2\xi'
+{\bf y}_c(\tilde\xi)\,,$$
$$
{\bf y}_c(\tilde\xi)=
\oint_{\partial D}{\partial\over \partial {\xi'}^a} G(\tilde\xi,\xi')\, {\bf y}(\xi')\, \sqrt g g^{ab}\epsilon_{bc}\,d{\xi'}^c
$$
$Z$ cancels against the source-independent parts of the functional integral (including the functional determinants) giving
$$
{\bf B}({\bf x})
=\int {d^{3} k\over 32\pi^4\alpha'}\, d^2\xi\, \epsilon^{ab}{\partial\over\partial{
\bf j}^a}\times{\partial\over\partial{
\bf j}^b}\,e^{-S''}\Big |_{{\bf j}=0}\,,
$$
$$
2\pi\alpha'S''=-{1\over 2}
\left\{i{\bf k}
+{
\bf j}^a{\partial\over\partial\tilde\xi^a}\right\}\cdot\left\{i{\bf k}
+
{\bf j}^b{\partial\over\partial\xi^b}\right\}\,G(\xi,
\tilde\xi)\Big|_{\tilde\xi=\xi} 
$$
$$-\left\{i{\bf k}
+{
\bf j}^r{\partial\over\partial\xi^r}\right\}\cdot{\bf y}_c(\xi)
+i{\bf k}\cdot{\bf x}
$$
This involves the Green function and its derivatives at co-incident points. To make this well-defined we introduce a regulator, $\epsilon>0$, via a spectral decomposition. If $u_\lambda (\xi)$ is an eigenfunction of the Laplacian belonging to eigenvalue $\lambda$,
vanishing on $\partial D$ then we can choose it to be real and using the fact that $\lambda>0$ take the regulated Green function to be \cite{Polyakov}
$$
G(\xi,\tilde\xi)=\sum_\lambda u_\lambda (\xi) \, u_\lambda (\tilde\xi)\,{e^{-\epsilon\lambda}\over \lambda}\,.
$$
Let $\psi$ denote the value of this at coincident points.
$\psi$ is greater than or equal to zero, vanishing only on the boundary $\partial D$. 
Furthermore, because $G$ is symmetric we have that 
$$
{\partial\over\partial\xi}\,G(\xi,
\tilde\xi)\Big|_{\tilde\xi=\xi} ={1\over 2}{\partial\over\partial\xi^a}\,\psi(\xi)
$$

\newpage

\noindent
enabling us to write $S''$ as
$$
2\pi\alpha'S''={1\over 2}{\bf k}^2\,\psi(\xi)
-{1\over 2}\,i{\bf k}\cdot {
\bf j}^a{\partial\over\partial{\xi}^a}\,\psi(\xi)
-{1\over 2}\,{\bf j}^a\cdot{\bf j}^b
{\partial^2\over\partial{\xi}^a_1\partial{\xi}^b_2}
\,G(\xi,
\tilde\xi)\Big|_{\tilde\xi=\xi} 
$$
$$-\left\{i{\bf k}
+{
\bf j}^r{\partial\over\partial\xi^r_1}\right\}\cdot{\bf y}_c
+i{\bf k}\cdot{\bf x}
$$
It is now straightforward to integrate over $\bf k$ in ${\bf B}({\bf x})$ and differentiate with respect to ${\bf j}$ to get
$$
{\bf B}({\bf x})=\int_D {d^2\xi\over2(4\pi^2\alpha'\psi)^{3/2}}\,\epsilon^{ab}\left({\partial 
{\bf y}_c\over\partial{\xi}^a}
\times({\bf y}_c-{\bf x})\,{1\over\psi}
{\partial \psi\over\partial{\xi}^b}+{\partial 
{\bf y}_c\over\partial{\xi}^a}\times{\partial 
{\bf y}_c\over\partial{\xi}^b}\right)\,
e^{-{({\bf y}_c-{\bf x})^2/(4\pi\alpha'\psi)}}\,.
$$
This splits into two integrals. In the first we change variables from $({\xi}^1,{\xi}^2)$ to $({\xi}^1,\eta=4\pi\alpha'\psi)$, supposing that on the boundary, $\partial D$, ${\xi}^2$ is constant. The form
of $\psi$ may be found by relating it to the heat-kernel
$$
{\cal G}(\xi,\tilde\xi,\tau)=\sum_\lambda u_\lambda (\xi) \, u_\lambda (\tilde\xi)\,e^{-\tau\lambda}\,,
\quad \psi=\int_\epsilon^\infty d\tau\,{\cal G}(\xi,\xi,\tau)\,,
$$
and then using the modification of the Seeley-de Witt expansion proposed in 
\cite{McAvity:1991xf}:
$$
{\cal G}={1\over 4\pi\tau}\sum_r \exp\left(
-{\sigma_r(\xi,\tilde\xi)\over 2\tau}\right)\Omega_r(\xi,\tilde\xi,\tau )\,,
$$
where the sum runs over all geodesic paths linking 
$\xi$ and $\tilde\xi$, including reflections at the boundary, and $\sigma(\xi,\tilde\xi)$ is twice the square of the path-length. For $\xi=\tilde\xi$ and small $\tau$ the path of zero length dominates for points away from the boundary, and $\Omega\sim 1+O(\tau)$. For points close to the boundary the 
shortest reflected path is also important, and the boundary conditions require that for this $\Omega\sim -1+O(\tau)$. So, if $\sigma/2$ is the square of the closest distance to the boundary then 
$$
\psi \sim \int_1^\infty{dt \over 4\pi t}  \left(1-e^{-\sigma /(2t\epsilon)}\right)\,,
$$
giving
$$
\psi \sim \sigma /(8\pi\epsilon),\quad{\rm for} \,\,\,\, 
\sigma<<\epsilon\,,
$$
whereas for $\sigma>>\epsilon$
$$
\psi \sim {\log (\sigma/\epsilon)/(4\pi)}\,.
$$
Consequently, as ${\xi}$ moves away from the boundary
$\eta$ varies from $0$ 
to a large positive value over a distance of order 
$\sqrt \epsilon$. In the interior of the domain this large value suppresses the integrand, consequently we only need to consider contributions to the integral from points close to the boundary. As the regulator is removed,
i.e. $\epsilon\rightarrow 0$, we can ignore the variation of ${\bf y}_c$
with $\eta$, replacing it with its boundary value, ${\bf y}$, and also take the $\eta$-integration limits to be $0$ and $\infty$.
So as the cut-off is removed
this integral becomes
\be
\int_0^\infty{d\eta\over \eta^{5/2}}\,\int {d{\xi}^1\over 2\pi^{3/2}}\,
{d 
{\bf y}\over d{\xi}^1}
\times({\bf y}-{\bf x})
\,
e^{-{({\bf y}-{\bf x})^2/\eta}}={1 \over 4\pi}\oint_C {d{\bf y}\times(
{\bf x}-{\bf y})\over ||{\bf x}-{\bf y}||^3}\,,
\label{firstint}
\ee
which is the Biot-Savart law. We now argue that the second integral 
\be\int_D {d^2\xi\over2(4\pi^2\alpha'\psi)^{3/2}}\,\epsilon^{ab}\,{\partial 
{\bf y}_c\over\partial{\xi}^a}\times{\partial 
{\bf y}_c\over\partial{\xi}^b}\,
e^{-{({\bf y}_c-{\bf x})^2/(4\pi\alpha'\psi)}}\,,
\label{secondint}
\ee
vanishes as the cut-off is removed. Again we ignore contributions to the integral from the interior of the domain
restricting our attention to a strip bordering the boundary of $D$ and change variables to $(\xi^1,\eta)$ so that (\ref{secondint}) becomes
\be
\int {d\xi^1 \,d\eta\over \eta^{3/2}}\,{\partial 
{\bf y}_c\over\partial{\xi}^1}\times{\partial 
{\bf y}_c\over\partial{\sigma}}{\partial 
{\bf \sigma}\over\partial\eta}\,e^{-{({\bf y}_c-{\bf x})^2/\eta}}\,,
\ee
Replacing ${\bf y}_c$ by its boundary value as before
gives
\be
\int d\xi^1 \left ({d 
{\bf y}\over d{\xi}^1}\times{\partial 
{\bf y}_c\over\partial{\sigma}}\int_0^{h}
{d\eta\over \eta^{3/2}}
{\partial 
{\bf \sigma}\over\partial\eta}e^{-{({\bf y}-{\bf x})^2/\eta}}\right)\,,
\ee
where $\eta$ ranges from $0$ to $h$ over the width of the strip.
Using $\psi \sim \sigma /(8\pi\epsilon)
$
gives
\bq
\int_0^{h}
{d\eta\over \eta^{3/2}}
{\partial 
{\bf \sigma}\over\partial\eta}e^{-{({\bf y}-{\bf x})^2/\eta}}
\sim 
{2\epsilon\over\alpha'}\int_0^{h}
{d\eta\over \eta^{3/2}}
e^{-{({\bf y}-{\bf x})^2/\eta}}\,.
\eq
This last integral is positive and less than
\bq
{2\epsilon\over\alpha'}\int_0^{\infty}
{d\eta\over \eta^{3/2}}
e^{-{({\bf y}-{\bf x})^2/\eta}}
={4\epsilon{\sqrt\pi}\over\alpha'|{\bf y}-{\bf x}|}\nonumber
\eq
which vanishes as the cut-off is removed.
Consequently 
only the first integral (\ref{firstint}) survives
and we have established that (\ref{BB}) satisfies the Biot-Savart law.
\be
\langle\,\int_\Sigma \delta^3({\bf x}-{\bf Y})
\,d{\bf S}({\bf Y}) \,\rangle_\Sigma={1\over 4\pi}\oint_C {d{\bf y}\times(
{\bf x}-{\bf y})\over ||{\bf x}-{\bf y}||^3}
=\nabla\times{1\over 4\pi}\oint_C {d{\bf y}\over ||{\bf x}-{\bf y}||}
\label{fin}
\ee

The result is independent of $g_{ab}$ so it is unchanged if we integrate over this metric degree of freedom as in Polyakov's approach to string theory \cite{Polyakov}. This independence is remarkable because
although (\ref{bprime}) itself does not contain 
$g_{ab}$ the computation of its expectation value required
a regulator and the use of $\psi$ which introduce such a dependence. String theory calculations are replete with 
quantities acquiring such `anomalous' dependence on the 
world-sheet metric, which make it difficult to average
over surfaces without imposing extra conditions, for example on the dimension of space or the mass spectrum 
of excitations. When the delta-function of (\ref{BB})
is represented as an integral over $k$ we are effectively summing over all the spectrum, and yet no mass-shell condition had to be imposed restricting this sum.

The choice of weight (\ref{avvS})
is a natural one to make in the context of string theory. It is also natural from the point of view of magnetostatics\footnote{I am grateful to Benjamin Doyon for pointing this out.} since the energy in the magnetic field (\ref{bprime}), ${1\over (2\mu_0)}\int d^3x \,{\bf B}^2$, is proportional to the area of $\Sigma$ albeit with a divergent coefficient and, as is well-known, the exponent in (\ref{avvS}) reduces to the surface area on eliminating $g_{ab}$ through its Euler-Lagrange equation.

\section{Time-dependence}

We now turn to our main problem which is the representation of the electromagnetic field of moving charges in terms of fluctuating lines of electric flux, so that the classical field results from a thermal average over a distribution of these lines. (We will treat the lines of force by assuming that they connect equal and opposite electrical charges and so have finite 
extent. Using this approach we could treat semi-infinite lines of force associated with single charges by taking the limiting case in which one charge in each pair is sent to infinity.)
If the charges of section 2 are now allowed to move then the electromagnetic field generated by the current-density
\be
J^\mu(x)=q\int^\infty_{-\infty} \delta^4( x-a)\, { \dot a}^\mu \, dt-q\int ^\infty_{-\infty}\delta^4( x-b) \,{ \dot b}^\mu \, dt
\label{curr}
\ee
will vary in time. Maxwell's formulation of electromagnetism is time-reversal invariant. The classical problem of computing this field by solving Maxwell's equations leads to solutions in terms of advanced or retarded potentials (or indeed linear combinations of the two) that have to be distinguished from each other by an application of common sense rather than from fundamental theory. We want to show that if instead the field is generated by an ensemble of lines of force in thermodynamic equilibrium the retarded solution arises naturally.

Assembling $\bf E$ and $\bf B$ into the antisymmetric tensor
$F_{\mu\nu}$ in the usual way by taking $F_{0i}=\epsilon_0\,E_i$ and
$\mu_0\,F_{ij}=-\epsilon_{ijk}\,B_k$ allows the two Maxwell equations containing sources to be written as 
\be
\partial^\mu \, F_{\mu\nu}=J_\nu \,,
\ee
which can be solved by taking $F_{\mu\nu}$ to be
\be
-q\int_\Sigma \delta^4(x-Y)\, d\Sigma_{\mu\nu}(Y)\,.
\label{Fsoln}
\ee
for any surface $\Sigma$ that spans the world-lines of the two charges. This solution describes a field supported on the surface $\Sigma$ which can therefore be interpreted as the world-sheet of a line of electric flux. To obtain a field that satisfies the remaining Maxwell equations
\be
\partial_\mu\,F_{\nu\rho}+\partial_\nu\,F_{\rho\mu}
+\partial_\rho\,F_{\mu\nu}=0\,.\label{covn}
\ee
we look for a suitable Boltzmann weight with which to construct an average over $\Sigma$:
\be
F_{\mu\nu}(x)=
-\langle q\int_\Sigma \delta^4(x-Y)\, d\Sigma_{\mu\nu}(Y)\rangle _{\Sigma}\,.
\label{Fsolnn}
\ee
As in section two we interpret this as an average over a distribution of many elementary strings, each associated with basic charge $q_0$ and in thermal equilibrium. $q$ rather than $q_0$ enters (\ref{Fsolnn}) because the individual contributions of the flux-lines must be summed.
We will ultimately construct this average but first we digress briefly by considering the problem in four-dimensional Euclidean space as this provides a useful step towards constructing the full Minkowski space theory. Because of the similarity between
(\ref{current}) and (\ref{curr}) this problem is solved by a
straightforward generalisation of the computation of the previous section. This can be done in any dimension, but if
we specialise to four dimensions then by taking the generalisation of (\ref{avvS}) to be
\be
\langle \Omega\rangle_\Sigma={1\over Z}\int {\cal D} {\bf Y}\,\Omega\,\exp\left({
-{1\over 4\pi\alpha'}\int_D g^{ab}\,G_{\mu\nu}\,{\partial{ Y^\mu}\over\partial\xi^a}\cdot{\partial{ Y^\nu}\over\partial\xi^b}\,\sqrt {g}\,d^2\xi}\right)
\,,\label{avvS'}
\ee
with $G_{\mu\nu}=\delta_{\mu\nu}$ we obtain as the generalisation of (\ref{fin})
\bq
4\pi^2\langle  \int_\Sigma \delta^4(x-Y)\, d\Sigma_{\mu\nu}(Y)\rangle _{\Sigma}
&=&\partial_\mu \left(\int {da_\nu \over ||x-a||^2}-\int {db_\nu \over||x-b||^2}\right)\nonumber\\
&&-\partial_\nu \left(\int {da_\mu \over ||x-a||^2}-\int {db_\mu \over||x-b||^2}\right)\nonumber\\\
\label{sol}
\eq
in which $1/||x-a||^2$ is a Euclidean Green function for the Laplacian.

This result can be Wick rotated to Minkowski space by $x^4\rightarrow -ix^0$ and using an $\epsilon$ prescription to encode the position of the poles. The effect is to replace the Green function
\be
{1\over ||x-a||^2}\rightarrow
{1\over ({\bf x}-{\bf a})^2-(x^0-a^0)^2+i\epsilon }
\label{Feyn}
\ee
This does give a solution to Maxwell's equations in Minkowski space, however because it was obtained by Wick rotating the functional integral it has the physical interpretation of being a quantum expectation value (and therefore relevant to Dirac's microscopic theory \cite{Dirac}) and not a thermal average. As a consequence the 
causal properties of this solution are not those we seek, but rather they are those inherited from the Feynman propagator
(\ref{Feyn}).
To construct the average we want we consider the general problem of constructing thermal Green functions in quantum theory. (It is not sufficient to consider classical statistical mechanics because strings are extended objects described by two-dimensional field theory on the world-sheet consequently this computation would be afflicted by the ultra-violet catastrophe unless we invoke quantum mechanics). In the quantum theory of a dynamical variable $\hat\varphi$ with Hamiltonian operator $\hat H$, associated eigenkets $|E\rangle$, and a set of time-dependent operators $\hat\Omega_1 (t_1), .., \hat\Omega_n(t_n)$ the finite-temperature Green functions are thermal averages of time-ordered products:
\be
\langle \hat\Omega_n(t_n)..\hat\Omega_1 (t_1)\rangle_T=
N\sum_E e^{-\beta E}\langle E|\hat\Omega_n(t_n)..\hat\Omega_1 (t_1)|E\rangle
\ee
with
$N=1/\sum_E e^{-\beta E}$ and $t_n>..>t_1$. Making the time
dependence of the operators explicit
\be
\hat\Omega_j(t_j)=e^{i(t_j-t_0)\hat H}\hat\Omega_j(t_0)e^{-i(t_j-t_0)\hat H}
\ee
and expressing the sum over energy eigenstates as a trace puts this into the form
\be
N\,{\rm Tr}\left (e^{-\beta \hat H}e^{i(t_{n+1}-t_0)\hat H}e^{-i(t_{n+1}-t_n)\hat H}\hat\Omega_n(t_0)e^{-i(t_n-t_{n-1})\hat H}..\hat\Omega_1(t_0)e^{-i(t_1-t_0)\hat H}\right)
\ee
where we have introduced a smallest time $t_0$ and a greatest time $t_n$.
The functional integral representation can be constructed in the usual way to give
\be
{1\over Z}\int {\cal D}\varphi \,e^{i\int_C L\,dt}\,\Omega_n(t_n)..
\Omega_1(t_1)
\ee
where $L$ is the Lagrangian related to $H$.
The contour $C$ consists of three straight-line segements. The first, $C_1$ runs just below the real axis from $t_0$ to $t_{n+1}-i\epsilon'$. The second,  $C_1$ runs from 
$t_{n+1}-i\epsilon'$ to $t_0-2i\epsilon'$ and the final segment, $C_3$ runs vertically down from $t_0-2i\epsilon'$
to $t_0-i\beta$. $\epsilon'$ is included to ensure convergence in the exponential factors. Note that the $\Omega$ are inserted only on the first segment, $C_1$. The trace is computed by identifying the values of the integration variable $\varphi$ at $t_0$ and 
$t_0-i\beta$. (See \cite{Niemi:1983nf} and references therein for a discussion of complex time contours and thermal Green functions).

We will interpret this construction as specifying the target space for our finite temperature theory. The world-sheets of the electric flux lines should wrap around the contour $C$.
We will take $\beta$ to be small (on the scale of the typical distances $|{\bf x -a}|$ and $|{\bf x -b}|$ involved) so that
we can neglect the contribution to that part of the world-sheet on $C_3$.
Ultimately we should send $x_0\rightarrow -\infty$ and 
$x_{n+1}\rightarrow \infty$ resulting in two infinite segements, $C_1$ and $C_2$. The edges of the sheet are the world-lines of the two charges, and these are duplicated on the two segments, however the operator whose Green function we are computing is restricted to $C_1$. There are thus two contributions to the thermal average. The first comes from the part of the world-sheet on $C_1$ and gives the Minkowski space version of (\ref{sol}) based on the Green function (\ref{Feyn}). The second comes from the part of the world-sheet on $C_2$. Because $x^0\in C_1$ but $a^0\in C_2$ the Green function is 
\be
{1\over ({\bf x}-{\bf a})^2-(x^0-a^0+i\epsilon'')^2}
\label{Feyn'}
\ee
instead of (\ref{Feyn}). ($\epsilon''$ is positive, proportional to $\epsilon'$ and has an irrelevant dependence on $x^0$ and $a^0$). When the direction in which $C_2$ is traversed is taken into account
these two contributions combine to give an expression like  
(\ref{sol}) but with the replacement
\bq
&&
{1\over ||x-a||^2}\rightarrow\nonumber\\
&&
{1\over ({\bf x}-{\bf a})^2-(x^0-a^0)^2+i\epsilon }
-{1\over ({\bf x}-{\bf a})^2-(x^0-a^0)^2 -2i(x^0-a^0)   \epsilon''}\,.\nonumber\\
\label{Feynm}
\eq
Now this is a representation of the retarded Green function
$$-2\pi i \theta(x^0-a^0)\,\delta\left(
(x^0-a^0)^2- ({\bf x}-{\bf a})^2\right)
=-{\pi i\delta\left(
x^0-a^0- |{\bf x}-{\bf a}|\right)\over |{\bf x}-{\bf a}|}
$$ 
Taking into account the factor of $i$  that $\delta^4(x-Y)$
acquires under Wick rotation
we find that the contribution to the Minkowski space thermal average 
\be
-\langle  {}\int_\Sigma \delta^4(x-Y)\, d\Sigma_{\mu\nu}(Y)\rangle _{\Sigma}\label{obj}
\ee
coming from the edges of the world-sheet that are the world-lines of the charges is
\bq
&&\partial_\mu \left(\int {da_\nu }{\delta\left(x^0-a^0-|{\bf x-a}|\right)\over 4\pi |{\bf x-a}|}-
\int {db_\nu }{\delta\left(x^0-b^0-|{\bf x-b}|\right)\over 4\pi |{\bf x-b}|}\right)
\nonumber\\
&&
-\partial_\nu \left(\int {da_\mu }{\delta\left(x^0-a^0-|{\bf x-a}|\right)\over 4\pi |{\bf x-a}|}-
\int {db_\mu }{\delta\left(x^0-b^0-|{\bf x-b}|\right)\over 4\pi |{\bf x-b}|}\right)\,.
\nonumber\\\
\label{retsol}
\eq
Since we are taking $\beta$ small so as to be able to ignore 
$C_3$ the trace is evaluated by identifying the string configurations at the end of $C_1$ and $C_2$ at times $t_0$ and $t_0-2i\epsilon'$ respectively, and then integrating over them. Before the integration is done the configuration of the string constitutes two fixed edges of the world-sheet described by two copies of the same curve displaced through $2i\epsilon'$ in time. Taking the edge attached to the end of $C_1$ to be  $Y^\mu=c^\mu$, say, there is a contribution to (\ref{obj}) of 
\be
\partial_\mu \left(\int {dc_\nu }{\delta\left(x^0-c^0-|{\bf x-c}|\right)\over 4\pi |{\bf x-c}|}\right)
-\partial_\nu \left(\int {dc_\mu }{\delta\left(x^0-c^0-|{\bf x-c}|\right)\over 4\pi |{\bf x-c}|}\right)\,.
\label{contribu}
\ee
Because this is supported on null-rays through  $Y^\mu=c^\mu$ it vanishes for fixed $\bf x$ when we send $t_0\rightarrow -\infty$ (provided that the world-lines of the two charges, 
$a^\mu$ and $b^\mu$ are not themselves null). Consequently only (\ref{retsol}) survives in this limit. Summing the contributions from all the elementary charges $q_0$
by simply multiplying this by $q$ results in the retarded solution to the Minkowski space Maxwell equations, $F_{\mu\nu}^{\rm ret}$, so we arrive at the conclusion that the retarded solution arises naturally from the thermal average:
\be
F_{\mu\nu}^{\rm ret}=-\langle  {q}\int_\Sigma \delta^4(x-Y)\, d\Sigma_{\mu\nu}(Y)\rangle _{\Sigma}
\ee
in the limit of small $\beta$. Notice that as in section two we have taken the temperature to be large.

\section{Conclusions}

The search for a unified description of physical phenomena is an old one. Faraday's idea that the basic atoms of such a description are lines of force may appear as no more than a  quaint notion that was rapidly side-lined by Maxwell's quantitative description of electromagnetism in terms of fields. However, from a modern perspective informed by string theory, his notion seems remarkably prescient, as indeed do Dirac's related ideas of over half a century ago.

We have used some of the world-sheet technology of string theory to show that classical electromagnetism can be interpreted as a consequence of the statistical mechanics of lines of force. This provides an underlying stochastic description of what is usually taken to be a deterministic problem encoded in the partial differential equations of Maxwell's theory. The result of this is that the retarded solution arises naturally, breaking the time-reversal invariance of the classical theory as a consequence of invoking thermodynamics. The direction of time is picked out because when we invoke the statistical mechanics of flux-lines we necessarily have to use quantum theory to avoid the ultra-violet catastrophe associated with the statistical mechanics of fields, since the world-sheets of strings are effectively field theories with one space and one time dimension. Quantum theory requires a Hamiltonian operator with a spectrum that is bounded from below. This allowed us to introduce $\epsilon$ insertions as convergence factors in our discussion of time dependence in section four, and it is these that pick out the retarded, rather than the advanced, solution to Maxwell's equations.

Electromagnetism, and gauge theories generally, are intimately connected with string theory and arise in a number of ways so we should consider how our computations are related to these standard approaches.

Conventionally photon
vertex operators are open-string operators inserted on the world-sheet boundary and correspond to coupling to an external $A_\mu$ field by adding to the action $\oint A_\mu\, dy^\mu$ so that $A_\mu$ acts as a source for the operator. 
The scattering amplitudes of photons of definite momenta $k^\mu$ and polarisation $E_\mu$ are thus obtained from the expectation values of vertex operators $q\oint e^{i k\cdot y}E\cdot d y$ on the world-sheet boundary, and the mass-shell condition $k^2= 0$ results from the requirement of the decoupling of the scale of the world-sheet metric.
In our flux string picture the electromagnetic field strength $F_{\mu\nu}$ is represented by a closed-string vertex operator that is inserted on the interior of the world-sheet and corresponds to coupling to an external $B^{\mu\nu}$ field by adding $\int qB^{\mu\nu}\,d\Sigma _{\mu\nu}$ to the action.
We can make a connection between the conventional open string vertex operators and our closed string ones by recalling the LSZ formalism in QED. This gives scattering amplitudes as time-ordered expectation values of on-shell field operators
\bq
\int d^4x \,e^{i k\cdot x}E^\mu\partial^2 A_\mu (x)
=-\int d^4x\, e^{i k\cdot x}E^\mu\partial^\nu F_{\mu\nu}(x)
\eq
since in QED $F_{\mu\nu}=\partial_\mu A_\nu-\partial_\nu A_\mu$ is an identity (although it only holds on average in our work), and $k\cdot E=0$. If we were to calculate this in terms of the flux string picture using (\ref{Fsolnn}) we would calculate the expectation value of 
\bq
q\int d^4x\, e^{i k\cdot x}E^\mu\partial^\nu 
\int_\Sigma \delta^4(x-Y)\, d\Sigma_{\mu\nu}(Y)
=q\oint_ {\partial\Sigma}e^{i k\cdot y}E\cdot d y
\eq
which is the usual open string insertion.

The averages over the flux string configurations that we have needed to construct $F_{\mu\nu}$ can be computed in any dimension because the world-sheet metric decouples from the calculation, consequently there is no mass-shell condition and we are free to use this off-shell $\delta$-function insertion. So our calculation is essentially one in non-critical string theory.

Gauge theories also arise from strings stretching between coincident D-branes. Our flux-lines stretch between the world-lines of electric charges which may be considered as D0-branes, however they are not coincident so that the flux-lines would not be associated with massless excitations. 

There is a further way in which our flux strings differ from usual string theory, and that is in their interactions. We have computed the classical electromagnetic field from the statistical mechanics of large numbers of flux-lines, but we could instead use this approach to pursue further Dirac's idea of building Quantum Electrodynamics from elementary flux lines. In the Euclidean functional integral approach to the quantum theory of the electromagnetic field coupled to charged particles we need to compute the expectation value of operators by integrating over the gauge field
\be
\langle\,\hat\Omega_1..\hat\Omega_n\,\rangle_{A}
={1\over Z_A}\int{\cal D}A\,e^{-{1\over 4}\int F^{\mu\nu}F_{\mu\nu}\,d^4x}\, \Omega_1(A)..\Omega_n(A)\,.
\ee
It is sufficient to take the operators to be Wilson loops,\cite{Strassler:1992zr},
i.e. 
$$\Omega_j=\exp -iq_j\oint_{C_j} A_\mu \,dx^\mu\,,$$ 
for arbitrary $C_j$. We will not address the dynamics of the charges, which requires integrating over the $C_j$, but 
focus on eliminating $A_\mu$.
The integral over the gauge-potential can be computed exactly as a functional of the curves $C_j$, because the exponent is quadratic in $A_\mu$, and the result is
$$
\exp -{1\over 2}\sum_{j} iq_j \int_{\Sigma_i} F_{\mu\nu}^c\,d\Sigma^{\mu\nu}_j\,,
$$
where $F_{\mu\nu}^c$ is the classical (Euclidean) electromagnetic field 
generated by the charge density $
\sum iq_j\oint_{C_j} \delta^4 (x-y)\, dy^\mu\,,
$ and $\Sigma_j$ is any surface spanning $C_j$.
We now represent the classical electromagnetic field as 
an average over flux-strings to obtain 
\be
\langle\,\hat\Omega_1..\hat\Omega_n\,\rangle_{A}
=\exp \sum_{i,j} {q_i q_j\over 2}\,\langle\,\int d\Sigma(X)_{\mu\nu}^i
\,\delta^4(X-Y)\,d\Sigma(Y)^{\mu\nu}_j\,\rangle_{\Sigma_i,\Sigma_j}
\ee
where we have introduced an extra averaging over the $\Sigma_i$ to obtain a more symmetrical result. Although this is not a complete theory it is clear that the basic interaction between the flux strings is a contact interaction rather than the splitting and joining interaction of conventional open string theory.


\acknowledgments
The author is grateful to Benjamin Doyon for stimulating discussions and to STFC for support under the rolling grant ST/G000433/1.

\end{document}